\documentstyle[12pt,aaspp4]{article}
\begin{document}

\def\newpage{\vfill\eject}
\def\vs{\vskip 0.2truein}
\def\pp{\parshape 2 0.0truecm 16.25truecm 2truecm 14.25truecm}
\def\fun#1#2{\lower3.6pt\vbox{\baselineskip0pt\lineskip.9pt
  \ialign{$\mathsurround=0pt#1\hfil##\hfil$\crcr#2\crcr\sim\crcr}}}
\def\core{{\rm core}}
\def\min{{\rm min}}
\def\max{{\rm max}}
\def\kpc{{\rm kpc}}
\def\esc{{\rm esc}}
\def\crit{{\rm crit}}
\def\pc{{\rm pc}}
\def\kms{{\rm km}\,{\rm s}^{-1}}
\def\cbh{{\rm cbh}}
\def\bh{{\rm bh}}
\def\df{{\rm df}}
\def\bulge{{\rm bulge}}

\lefthead{Miralda-Escud\'e \& Gould}
\righthead{High Velocity Star Formation in the LMC}
%%%%%%%%%%%%%%%%%%%%%%%
%%%%%%%electronic submission format
\title{A Cluster of Black Holes at the Galactic Center}
\author{Jordi Miralda-Escud\'e$^1$ and Andrew Gould}
\affil{Department of Astronomy, The Ohio State University,
Columbus, OH 43210, USA}
\affil{{}$^1$ Alfred P. Sloan Fellow}
\authoremail{jordi@astronomy.ohio-state.edu, gould@astronomy.ohio-state.edu}
%%%%%%%%%%%%%%%%%%%%%%%%%%%%%%%

\begin{abstract}

If the stellar population of the bulge contains black holes formed in the final
core collapse of ordinary stars with $M \ga 30 M_{\odot}$, then about 25,000
stellar mass black holes should have migrated by dynamical friction into the
central parsec of the Milky Way, forming a black hole cluster around the
central supermassive black hole. These black holes can be captured by the
central black hole when they randomly reach a highly eccentric orbit due to
relaxation, either by direct capture (when their Newtonian peribothron is less
than 4 Schwarzschild radii), or after losing orbital energy through
gravitational waves. The overall depletion timescale is $\sim 30$ Gyr, so most
of the 25,000 black holes remain in the central cluster today. The presence of
this black hole cluster would have several observable consequences. First, the
low-mass, old stellar population should have been expelled from the region
occupied by the black hole cluster due to relaxation, implying a core in the
profile of solar-mass red giants with a radius of $\sim 2\,\pc$ (i.e., $1'$).
The observed central density cusp (which has a core radius of only a few arc
seconds) should be composed primarily of young ($\la 1\,$Gyr) stars. Second,
flares from stars being captured by supermassive black holes in other galaxies
should be rarer than usually expected because the older stars will have been
expelled from the central regions by the black hole clusters of those galaxies.
Third, the young ($\la 2$ Gyr) stars found at distances $\sim 3 - 10\, \pc$
from the Galactic center should be preferentially on highly eccentric orbits.
Fourth, if future high-resolution $K$-band images reveal sources microlensed by
the Milky Way's central black hole, then the cluster black holes could give
rise to secondary (``planet-like'') perturbations on the main event.

\end{abstract}

\keywords{black hole physics -- 
Galaxy: center -- Galaxy: kinematics and dynamics}
%%%%%%%%%%%%%%%%%%%%%%%%%%%%

\setcounter{footnote}{0}
\renewcommand{\thefootnote}{\arabic{footnote}}

\section{Introduction}

  The measurement of proper motions and radial velocities of stars
within the central parsec of the Galaxy has led to the conclusion that
a black hole of mass $3.0\pm 0.3\times 10^6 M_\odot$ is present in the
center (Eckart \& Genzel 1997; Genzel et al.\ 2000).
There is also increasing evidence that massive black holes are found in
the centers of other galaxies (Richstone et al.\ 1998).

  The central region of the Galaxy is also peculiar because the
relaxation time among stars can be shorter than the age of the Galaxy,
owing to the
high density. The process of relaxation leads to a stellar cusp, which
has a density profile $\rho \propto r^{-7/4}$ when all the stars have the
same mass (Bahcall \& Wolf 1976). Several interesting physical processes
take place among the stars in this cusp: stars can come close enough to
physically collide with each other, and they can also come sufficiently close
to the black hole to be tidally disrupted or swallowed
(e.g., Frank \& Rees 1976; Lightman \&
Shapiro 1977; Quinlan, Hernquist, \& Sigurdsson 1995; Sigurdsson \& Rees
1997).

  One of the consequences of the relaxation is that the most massive
objects will sink to the center of the stellar cusp. Among an old stellar
population, the most massive objects should be black holes formed in the
final core collapse of massive stars. We assume in this paper that most
massive stars with $M \ga 30\, M_\odot$ produce black holes, most with a
mass of $7\, M_\odot$ (Bailyn et al.\ 1998). The high mass of
these black holes implies that their dynamical friction time to move to
the center of the Galaxy is shorter than a Hubble time over a much larger
volume than the one where ordinary stars have a short relaxation time.
We will find in \S 2 that this should lead to the formation of a cluster
of stellar black holes around the central supermassive black hole
(hereafter ``Sgr A*''), and
that other stars are ejected from the region occupied by this cluster.
In \S 3 we discuss the rate at which the black holes in this
cluster are captured by Sgr A*, and we find that most of the
black holes should still be present in the cluster.   Several observable
consequences of the presence of this black hole cluster are discussed in
\S 4.

\section{Cluster Formation}

	The deprojected light profile in the inner kpc of the Galaxy scales as
$r^{-1.8}$, while the predicted profile around a massive black hole 
$(r\la1\,\pc)$ scales as $r^{-7/4}$ (Bahcall \& Wolf 1976).  For simplicity,
we therefore adopt a density profile $\rho(r)$
\begin{equation}
\rho(r) \propto r^{-7/4}.
\label{eqn:betanought}
\end{equation}
	From the model fit of Genzel et al.\ (2000) to the velocity dispersion
data, we find that the total mass within $r_0=1.8\,\pc$ is $2\,M_\cbh$
(see their Fig. 17), where
\begin{equation}
M_\cbh = 3\,\times 10^6\,M_\odot
\label{eqn:mcbh}
\end{equation}
is the mass of Sgr A*.
Hence, the total distributed mass inside 1.8 pc is $M_\cbh$, and
the density profile is 
\begin{equation}
\rho_*(r) = {5\over 16\pi}
\,{M_\cbh\over r_0^3}\,\biggl({r\over r_0}\biggr)^{-7/4},\qquad 
r_0\equiv 1.8\,\pc.
\label{eqn:rhor}
\end{equation}

	We assume that this density profile is entirely composed of stars,
brown dwarfs, and stellar remnants.  To calculate the mass fraction of 
black holes, $\eta_\bh$, we use the following initial mass function.
For the range $M>1\,M_\odot$, we adopt a Salpeter law $d N/d m\propto 
m^{-\alpha}$ with $\alpha=2.35$.  For $0.7\,M_\odot<m<1\,M_\odot$, we adopt
$\alpha=2$ from Zoccali et al.\ (2000).  For $0.05\,M_\odot<m<0.7\,M_\odot$, 
we adopt $\alpha=1.65$ by correcting the result of Zoccali et al.\ (2000)
for binaries according to the adjustment of Gould, Bahcall, \& Flynn (1997)
and by extending the power law beyond the last observed point at
$0.15\,M_\odot$.  We cut off the mass function at $m\sim 0.05\,M_\odot$
in accordance with the prelimary indications from microlensing (Han \& Gould
1976).  We assume that all progenitors with masses 1--8 $M_\odot$ have 
become $0.6\,M_\odot$ white dwarfs, those with masses 8--30 $M_\odot$ have 
become $1.4\,M_\odot$ neutron stars, and those with masses 30--100 
$M_\odot$ have become $7\,M_\odot$ black holes.  We then find 
\begin{equation}
\eta_\bh=1.6\%,\qquad \langle m\rangle = 0.23\,M_\odot,
\label{eqn:etabh}
\end{equation}
where $ \langle m\rangle$ is the mean mass of the population.

	Once this population is formed, the black holes will sink toward
the center on the dynamical friction timescale (Binney \& Tremaine 1987)
\begin{equation}
t_\df^{-1} = \ln\Lambda {4\pi G\rho G m_\bh\over v^3} \,
\int_0^v d u\,4\pi u^2 f(u) ~,
%H\biggl({v^2\over 2\sigma^2}\biggr),
%\qquad H(x)\equiv {2\over \pi^{1/2}}\int_0^x d y\, y^{1/2}e^{-y},
\label{eqn:tdf}
\end{equation}
where $m_\bh=7\,M_\odot$ is the mass of the black hole, $v$ is its velocity,
$f(u)$ is the velocity distribution of the ambient stars, and the integral
gives the fraction of ambient stars with speeds below $v$.
For a Gaussian velocity distribution with dispersion $\sigma$, and a
typical black hole speed $v^2=3\sigma^2$, the value of the integral is
$0.54$. For the Keplerian part of the potential ($r<r_0$), 
$\ln\Lambda = \ln(M_\cbh/m_\bh)=13$.  For $r>r_0$, ${\rm ln}\, \Lambda$
rises slightly but we ignore this in the interest of simplicity.
We evaluate $t_{\df,0}$, the fiducial dynamical friction time at $r_0$,
\begin{equation}
t_{\df,0} = 1.4\,\rm Gyr
\label{eqn:tdf0}
\end{equation}
and note its scaling in the two regimes
\begin{equation}
t_\df = t_{\df,0}\biggl({r\over r_0}\biggr)^{1/4}\quad (r<r_0),\qquad
t_\df = t_{\df,0}\biggl({r\over r_0}\biggr)^{17/8}\quad (r>r_0).
\label{eqn:tdfscale}
\end{equation}
Hence, after a time $t$, all the black holes that were 
originally within a radius $r$ will collect in a cluster near the center, 
where $r$ is given by,
\begin{equation}
{r\over r_0} = \biggl({t\over 4 t_{\df,0}}\biggr)^4\quad (t<4t_{\df,0}),
\qquad
{r\over r_0} = \biggl[{17\over 7}
\biggl({t\over t_{\df,0}}-4\biggr)+1\biggr]^{8/17}\quad (t>4t_{\df,0}).
\label{eqn:rlimit}
\end{equation}
This implies an infall rate of black holes
$$ {d N_\bh\over dt} = {5\, \eta_\bh\, M_\cbh\over 4\, t_{\df,0}\, m_\bh}
\biggl({t\over 4 t_{\df,0}}\biggr)^4\quad (t<4t_{\df,0}) ~, $$
\begin{equation}
{d N_\bh\over dt} = {10\, \eta_\bh\, M_\cbh\over 7\, t_{\df,0}\, m_\bh}
\biggl[{17\over 7}
\biggl({t\over t_{\df,0}}-4\biggr)+1\biggr]^{-7/17}\quad (t>4t_{\df,0}).
\label{eqn:dndt}
\end{equation}

	If we assume that the bulge formed at a time $t_\bulge\sim 10\,$Gyr,
then from equation (\ref{eqn:rlimit}), all the black holes within a radius
$ r_\df = 5\,\pc$ will have migrated to the center by now.
The cumulative total and current rate of
precipitation are therefore
\begin{equation}
N_\bh\sim 2.4\times 10^4,\qquad 
{d N_\bh\over d t} \sim 2.9\,\rm Myr^{-1}.
\label{eqn:numbh}
\end{equation}
In other words, provided that our assumption of the fraction of massive
stars that form black holes in their final core collapse is correct, we
must conclude that {\it a large number of stellar black holes, with a
total mass of $\sim 5\%$ of the Sgr A* mass, have migrated to
the center and, unless they have subsequently been captured by Sgr A*,
they should have formed a cluster of black holes in the
center of the stellar cusp}.

  As the black holes precipitate, they start dominating the total density
in some central region at some point, and then the low-mass stars are
expelled from this region over a relaxation time. Assuming that most of 
the energy is lost from the cluster by direct capture of black holes near
the center, the black holes should also relax to a density profile
proportional to $r^{-7/4}$, for which the outward energy flow is constant.
As this energy flow is transmitted to the low-mass stars outside
the black hole cluster, the cluster will need to expand and push out the
low-mass stars. 
To derive the relative density of the black hole profile
compared to the empirically normalized stellar profile, we invoke the
steady state energy-flow condition between two species of stars A and B
of mass $m_A$ and $m_B$, which dominate the total density $\rho_A$ and
$\rho_B$ at radii $r_A$ and $r_B$, respectively. The
total energy at radius $r$ is proportional to $\rho \sigma^2 r^3$, and
the relaxation time is proportional to $\sigma^3/(\rho m)$.
Therefore, the constant energy flow condition yields
\begin{equation}
{\rho_A \sigma_A^2 r_A^3\over \sigma^3/(\rho_A m_A)}=
{\rho_B \sigma_B^2 r_B^3\over \sigma^3/(\rho_B m_B)},
\label{eqn:equil}
\end{equation}
Making use of the Kepler-potential relation $\sigma^2\propto r$, this implies,
\begin{equation}
{\rho_A\over\rho_B} = 
\biggl({r_A\over r_B}\biggr)^{-7/4}
\biggl({m_A\over m_B}\biggr)^{-1/2}.
\label{eqn:rhoab}
\end{equation}
Thus, the mass density of black holes in the central region is {\it below}
that implied by extrapolating equation (\ref{eqn:rhor}), by a factor
$(\langle m\rangle/m_\bh)^{1/2}$, that is,
$\rho_\bh(r) = 0.18\rho_*(r)$. Hence, if all of the black holes
precipitated over the lifetime of the galaxy from a radius of $5\pc$
remain in the cluster at present, the
cluster should extend over a radius $r_\bh$,
\begin{equation}
r_{\bh} = \biggl({\eta_\bh^2 m_\bh\over\langle m\rangle}
\biggr)^{2/5}\,5\,\pc
= 0.7\,\pc.
\label{eqn:rbh}
\end{equation}
The timescale to achieve this expansion is the relaxation time in the
expanded (lower density) cluster, which is $\sim 3$ times longer than
the dynamical friction timescale because of the lower density.

\section{Rate of Capture of the Black Holes}

  In the previous section we found that about 24,000 stellar black holes
should have migrated to the center of the stellar cusp around Sgr A*.
We now address the question of the rate at which these black
holes will be removed from the cluster by coalescing with Sgr A*.
If this rate is low, most of the black holes should be in the
cluster at present. If the rate is high enough, then many fewer black
holes will be present, and a balance between the rate at which black holes
are precipitating in the cluster by dynamical friction and the rate at
which they are being captured should be established.

  The dominant process by which black holes will be eliminated is by a
random walk into a highly eccentric orbit as their orbits change over the
relaxation timescale, from which they can be captured by Sgr A*.
This process was first studied by Frank \& Rees (1976). In the case
of stars, tidal disruption can eliminate them from the cluster once they
come close enough to Sgr A*; obviously, orbiting black
holes will be eliminated only when they are swallowed by Sgr A*,
possibly after having lost orbital energy by emitting
gravitational waves.

  Before describing in more detail the mechanism by which black holes
are captured, we need to discuss the process of orbital diffusion by
which black holes will migrate into the eccentric orbits from which they
can be captured.

\subsection{Orbital Diffusion}

  A black hole can be captured by Sgr A* from an orbit of any semimajor
axis, provided that its peribothron $q=a(1-e)$ is sufficiently small.
This will lead to a distribution of black holes in phase space that is
strongly depleted at eccentricities very close to unity, and diffusion
of black holes will take place toward orbits of decreasing peribothron.
In order to investigate quantitatively this black hole migration,
we first evaluate the diffusion tensor in velocity space.
We sketch the derivation here and leave the details and
the justifications for the various approximations to Appendix A.

	The diffusion equation is given by
\begin{equation}
\nabla_v\cdot {\bf j} + {\partial f\over \partial t} = 0,\qquad
j_k \equiv -\sum_l\kappa_{kl}{\partial f({\bf v,r},t)\over \partial v_l}
\label{eqn:diffeq}
\end{equation}
where $f$ is the phase-space density, $\bf j$ is the ``velocity current
density'',  and $\kappa_{kl}$ is the diffusion tensor.  By symmetry,
$\kappa_{kl} =$ diag($\kappa_\perp,\kappa_\perp,\kappa_r$), where
$\kappa_\perp$ and $\kappa_r$ are the components of $\kappa$ perpendicular
and parallel to the radial direction. In general, the diffusion tensor
depends on the spatial position and the velocity. A useful physical
interpretation of the diffusion tensor components is that, over a
small interval of time $\delta t$, the rms change in the velocity of a
black hole in a direction perpendicular to its initial velocity is equal
to $(2\kappa_\perp\, \delta t)^{1/2}$. The total rms change in any direction
is therefore $(2\, tr(\kappa) \delta t)^{1/2}$ (where {\it tr} means the trace),
and the relaxation time is of order $v_\esc^2/[2 tr(\kappa)]$.

We assume that the
unperturbed (by black-hole capture) phase-space density,
$f_0$, is a function only of the energy, and hence find, for
a $\rho\propto r^{-\alpha}$ density profile in a Kepler potential,
\begin{equation}
f_0({\bf u,r}) = g\biggl({u^2\over v_\esc^2}\biggr)h(r),\qquad 
g(x) \equiv (1 - x)^{\alpha-3/2}
\Theta(1-x),
\label{eqn:aone}
\end{equation}
\begin{equation} 
h(r)\equiv
{(3-\alpha)\alpha!\over 8\pi^2(\alpha-3/2)!(1/2)!}
{N_\bh\over(2 G M_\cbh r_\bh)^{3/2}}
\biggl({r\over r_\bh}\biggr)^{3/2 - \alpha},
\label{eqn:atwo}
\end{equation}
where $v_\esc(r)$ is the local escape velocity, $N_\bh$ is the total number
of black holes within a radius $r_\bh$, $M_\cbh$ is the central mass, and
$\Theta$ is a step function.  Thus, for $\alpha=7/4$, 
$f_0\propto r^{-1/4}(1-v^2/v_\esc^2)^{1/4}$.
In Appendix A, we show that for $\alpha=3/2$, the velocity dependences of
the parallel and perpendicular components of the diffusion tensor are
\begin{equation}
\kappa_\perp(v,r) = \biggl(1 - {1\over 5}\,
{v^2\over v_\esc^2}\biggr)\kappa_0(r),\qquad
\kappa_r(v,r) = \biggl(1 - {3\over 5}\,{v^2\over v_\esc^2}\biggr)\kappa_0(r),
\label{eqn:cone}
\end{equation}
where $\kappa_0$ is the (isotropic) diffusion coefficient at $v=0$. We 
also justify in the Appendix using this velocity dependence as an
adequate approximation for the case $\alpha=7/4$, for which we find,
\begin{equation}
\kappa_0(r) = 
{7\over 6\sqrt{\pi}}\,{(3/4)!\over (1/4)!}\,
{N_\bh\, (G m_\bh)^2\, v_\esc^2\over(2 G M_\cbh r_\bh)^{3/2}} \,
\biggl({r\over r_\bh}\biggr)^{-1/4}\, \ln \Lambda_1 ~,
\label{eqn:aseven2}
\end{equation}
The quantity $\ln \Lambda_1$ can be slightly smaller than
$\ln \Lambda= \ln(M_{\cbh}/m_{\bh})$, as discussed in Appendix A,
depending on the
scale over which the diffusion takes place, because scatterings with large
velocity changes do not contribute to the diffusion rate over a small
range of velocities.

\subsection{Three Regimes of Capture}

  Before identifying the condition for capture from highly eccentric
orbits, it will be useful to compute the core radius, $r_c$, where
the energy-loss time to gravitational radiation on a circular orbit is
equal to the relaxation time. We define the relaxation time for circular
orbits as $t_{r,0} = 
v_c^2/[2 {tr}(\kappa)]$, where $v_c= (GM_\cbh/r)^{1/2}$ is
the circular velocity.
The energy loss by gravitational radiation for circular orbits is
\begin{equation}
{d \ln E\over d t}={64\over 5}{G^3m_\bh M_\cbh^2\over c^5 r_c^4} ~.
%\xi\biggl({r_c\over r_0}\biggr)^{1/4}t_{\df,0}=1.
\label{eqn:coreeq}
\end{equation}
Equating this to $t_{r,0}^{-1}$, we find,
$$
r_c = 1.57 \, \left( {G M_\cbh\over c^2}\right)^{2/3} \, r_\bh^{1/3}\,
\left( { M_\cbh \over N_\bh m_\bh \ln\Lambda } \right)^{4/15}  
$$
\begin{equation}
=
8.6\,{\rm AU}\, 
\biggl({M_\cbh \over 3\times 10^6\,M_\odot }\biggr)^{2/3}
\biggl({r_\bh\over 0.7\,\pc}\biggr)^{1/3}
\biggl({M_\cbh \over 1.38 N_\bh m_\bh \ln\Lambda } \biggr)^{4/15} ~.
\label{eqn:rceval}
\end{equation}

  This radius is extremely small compared to the cluster radius $r_{bh}$.
If black holes were captured by diffusing to orbits with $a\sim r_c$, and
then losing energy by gravitational waves at low eccentricity, the
capture rate would therefore also be extremely small. In reality, as
we shall see now, black holes will be captured from a large range of
radii, mostly from highly eccentric orbits.  We must therefore define
a relaxation time scale for these high eccentricity orbits.

	To do this, we first calculate the orbit-averaged rate of change
of the peribothron $q\equiv a(1-e)$ in a highly eccentric orbit.  From
angular momentum conservation, we have $v_\perp^2 =(1+e)q G M_\bh/r^2
\rightarrow q v_\esc^2/r$.  Hence, 
\begin{equation}
P{d\langle q\rangle\over dt} =
\int_0^P d t {2\kappa_\perp r\over v_e^2} =
2^{5/4}\ln\Lambda_1{53\over 33}\,
\biggl({m_\bh\over M_\cbh}\biggr)^2 N_\bh
\biggl({a\over r_\bh}\biggr)^{9/4}r_\bh \,,
\label{eqn:pdq}
\end{equation}
We then define the eccentric relaxation time $t_{r,1}(a) = 
(d\langle \ln q\rangle/dt)^{-1}$.
We now use $t_{r,1}$ to demonstrate that there are three regimes of capture:
1) for $a<a_{\rm trans}$, capture is dominated by gravitational radiation,
2) for $a_{\rm trans}<a<a_\crit$, it is dominated by direct capture, and
3) for $a>a_\crit$, capture falls off very rapidly and can be ignored.

  A particle orbiting around a Schwarzschild black hole with
semi-major axis $a$ much greater than the Schwarzschild radius 
will be directly captured by the black hole if its peribothron $q$,
computed by extrapolating the Newtonian orbit, is less than 4
Schwarzschild radii (e.g., Misner, Thorne, \& Wheeler 1973). When
$q=8GM_\cbh/c^2$, the particle is actually brought to 2
Schwarzschild radii by relativistic effects, where the maximum of the
effective radial potential is located; the particle has then overcome
the angular momentum barrier and can directly fall into the black
hole. Therefore, the phase space density should drop to zero below
the minimum peribothron
\begin{equation}
q_{\min} = {8GM_\cbh\over c^2 } \simeq 0.24\, {\rm AU} ~.
\label{eqn:direct}
\end{equation}

  The black hole can also be captured at larger peribothron if the
timescale to change the eccentricity by diffusion is longer than the
timescale for losing its orbital energy by gravitational waves.
For $1-e \ll 1$, the gravitational radiation decay rate is,
\begin{equation}
{d\ln E\over dt}= {170\over 3}\, (2q)^{-7/2}\,
{ G^3\, m_\bh\, M_\cbh^2\over c^5\, a^{1/2} } ~.
\label{eqn:greccen}
\end{equation}
We determine $q_\min$, the minimum peribothron
that avoids capture, by setting $d\ln E/dt = t_{r,1}^{-1}$,
\begin{equation}
q_\min(a) = 0.35 \,\biggl({8 G M_\cbh\over c^2}\biggr)\,
\biggl( {M_\cbh \over 1.73 N_\bh m_\bh \ln\Lambda_1 } \biggr)^{2/5} \,
\biggl( { r_\bh \over a } \biggr)^{1/2}\, ,
\label{eqn:direct2}
\end{equation}
where the factor $1.73$ is the ratio $M_\cbh/(N_\bh m_\bh
\ln\Lambda_1)$ for the values we use in this paper, and for
$\ln\Lambda_1=9.7$ (see Appendix A).
Hence, the transition from capture by gravitational radiation to direct
capture occurs at $a_{\rm trans}= 0.35^2 r_\bh$ = 17,000 AU.

  Above some critical semimajor axis $a_\crit$, the diffusion of $q$
over a single period exceeds $q_\min$ as given by equation
(\ref{eqn:direct}) for direct capture. Hence, during each period,
$P\sim a^{3/2}$, the black holes are captured with a probability that
decreases with semimajor axis as $(q_\min/a)\sim a^{-1}$, so that the
capture rate falls off $ \propto a^{-5/2}$. Since the mass within radius
$r$ increases as $r^{3-\alpha}$, captures from orbits with $a \gg a_\crit$
produce a negligible loss of black holes.
We evaluate $a_\crit$ by setting $t_{r,1}=P$, and find
\begin{equation}
a_\crit = 0.41\,{\rm pc}
\biggl({N_\bh\over 2.4\times 10^4}\biggr)^{-4/9}
\biggl({M_\cbh\over 3\times 10^6\,M_\odot}\biggr)^{4/3}
\biggl({m_\bh\over 7\,M_\odot}\biggr)^{-8/9}
\biggl({r_\bh\over 0.7\,\rm pc}\biggr)^{5/9}.
\label{eqn:rcrit}
\end{equation}

\subsection{Capture Rate From The Loss Cylinder}

	While $\kappa$ is a function of both $\bf v$ and $r$, we will solve 
the diffusion equation at fixed $r$, and temporarily assume that $\kappa$
is independent of $\bf v$. We will introduce variation
in $\kappa$ only when we evaluate the loss rate.  This is a very good
approximation, as we discuss in Appendix A.
We focus first on the case of $a_{\rm trans}<a<a_\crit$,
 for which capture is typically direct as described by equation
(\ref{eqn:direct}).  Since $q_\min$ is independent of $a$, the captured 
orbits at fixed $r$ are characterized by a cylinder in velocity space,
\begin{equation}
{v_\perp\over v_\esc}\leq \sqrt{q_\min\over r}=2{v_\esc\over c}
,\qquad {|v_r|\over v_\esc}\leq 1,
\label{eqn:cylind}
\end{equation}
where $v_r$ and ${\bf v}_\perp$ are the radial and perpendicular
components of the velocity. Note that this structure in velocity space
is definitely a ``loss cylinder'' and not a ``loss cone'' as it is often
described in the literature. Making use of
yet another good approximation described in Appendix A, we solve
the steady-state diffusion equation (\ref{eqn:diffeq}), at fixed $v_r$:
\begin{equation}
f({\bf v}_\perp;v_r) = f_0(v_r)\biggl[1 - {\ln(v_\perp^2/(v_\esc^2-v_r^2))
\over \ln(q_\min/r)}\biggr].
\label{eqn:diffsolv}
\end{equation}
Hence, the capture rate per unit volume, $dC/dV$, is given by
\begin{equation}
{dC\over dV} = -\int_{{\rm cyl-bnd}}{\bf j}\cdot d{\bf A}  =
8\pi\int_0^{v_\esc} d v_r\, {\kappa_\perp(v_r)f_0(v_r)\over\ln(r/q_\min)} ~,
\label{eqn:capone}
\end{equation}
where $d{\bf A}$ is the area element on the boundary of the capture
cylinder (``cyl-bnd'').  
We evaluate this using equations (\ref{eqn:aone}) and (\ref{eqn:cone}),
$$ {d C\over d V}
= {35\over 16\pi^{3/2}}\,{(3/4)!\over (1/4)!}\,{\ln\Lambda_1\over 
\ln(r/q_\min)}
\,{N^2_\bh(G m_\bh)^2 \over(2 G M_\cbh r_\bh)^{3/2}}
\biggl({r\over r_\bh}\biggr)^{-2}r_\bh^{-3} $$
\begin{equation}
 = {6\kappa_0\over v_\esc^2}\, {n(r)\over \ln(r/q_\min) } ~,
\label{eqn:captwo}
\end{equation}
where $n(r) = 5 N_\bh/(16\pi r_\bh^3)\, (r/r_\bh)^{-7/4}$
is the number density of black holes at radius $r$, and
$6 \kappa_0/v_\esc^2$ is of order the relaxation time.
Integrating over volume, the total capture rate within 
some maximum radius $r_\max$ 
(determined below) is
\begin{equation}
C=
k{\ln\Lambda_1\over\ln(r_\max/q_\min)}N_\bh^2
\biggl({m_\bh\over M_\cbh}\biggr)^2\,{r_\max\over r_\bh}\,{2\pi\over P_\bh}
%\begin{equation}
%= {2(3-\alpha)^2(\alpha-1)(2\alpha+1)(\alpha-1)!\over 3(9-4\alpha)
%(\alpha-1/2)!(1/2)!}
%{77\over 16\sqrt{\pi}}\,{(3/4)!\over (1/4)!}\,
%{N^2_\bh(G m_\bh)^2\over(2 G M_\cbh r_\bh)^{3/2}}
%\biggl({r_\max\over r_\bh}\biggr)%^{9/2 - 2\alpha}
%{\ln\Lambda_1\over \ln(1-e_\min)^{-1/2}},
\label{eqn:capthree}
\end{equation}
where $k=(2\pi)^{-1/2}(35/8)(3/4)!/(1/4)!\sim 1.76$, $P_\bh$ is the
orbital period at $a=r_\bh$, and 
where we have made the evaluation treating 
$\ln(r/q_\min)\rightarrow\ln(r_\max/q_\min)$ and
$\ln\Lambda_1$ as constants. Notice that the capture rate increases as
$N_\bh^{4/5}$ as black holes are added to the cluster, because
$r_\bh \propto N_\bh^{4/5}$, and $P_\bh \propto N_\bh^{6/5}$.

  As described already by Frank \& Rees (1976), the result we have
reached in equations (\ref{eqn:captwo}) and (\ref{eqn:capthree}) shows
that {\it the capture rate of the black holes is essentially
proportional to the relaxation time, and the fact that $q_\min/r \ll 1$
increases the time required for the
black hole to find a capture orbit only logarithmically}. The reason
is that, over a relaxation time, an orbiting black hole will
totally change its eccentricity not only due to some large deflection
in a close encounter, but also due to many small deflections that will
change the eccentricity by very small amounts, allowing the black hole
to effectively sweep over all possible eccentricities and find the
very narrow range of eccentricity where it can be captured. However,
this is no longer true for $a > a_\crit$: when $q$ is brought
below $q_\min$ by the random deflections, the black hole will most
likely miss being captured unless it happens to be at peribothron.

  The maximum radius $r_\max$ of integration of the capture rate
in equation (\ref{eqn:capthree}) must therefore be of order $a_\crit$.
For a highly eccentric orbit, the time-averaged radius is
$\langle r\rangle = (3/2)a$. We therefore adopt $r_\max = (3/2)a_\crit$.
Since capture is dominated by black holes near $r_\max$,
we evaluate $\ln(r/q_\min)$ there and find,
\begin{equation}
\ln(r_\max/q_\min) = \ln{3 c^2 a_\crit\over 16 G M_\cbh}
\simeq 13.2.
\label{eqn:rcrit2}
\end{equation}
At the same time, the term $\ln\Lambda_1$ can be approximated as
(see Appendix A)
\begin{equation}
\ln\Lambda_1 \simeq \ln\Lambda - {1\over 4}\ln{c^2 r_\max\over 8 G M_\cbh}
\simeq 9.7
\label{eqn:rcrit3}
\end{equation}
Hence, the ratio of logarithms in equation (\ref{eqn:capthree}) is
$\ln\Lambda_1/\ln(r_\max/q_\min)\sim 0.73$.
We are finally able to evaluate the capture rate explicitly,
\begin{equation}
C \simeq {N_\bh\over 30\,\rm Gyr}.
\label{eqn:capfour}
\end{equation}
Since this timescale is much longer than a Hubble time, most of the 24,000 
black holes that have entered the cluster are still in it and have not been
captured.  Hence, the actual radius of the black hole cluster is close to
our initial estimate given by equation (\ref{eqn:rbh}).

	Note that we have everywhere used the direct capture formula 
(\ref{eqn:direct}) to calculate $q_\min$ rather than the gravitational
radiation formula (\ref{eqn:direct2}), which applies at
$r<a_{\rm trans}\sim 17,000\,$AU.  Recall, however, that
$q_\min$ only enters into the logarithm term.  The capture rate from
the inner part of the cluster is therefore higher than we have assumed,
but not dramatically. Notice that since $q_\min$ is a function of
$a$ in equation  (\ref{eqn:direct2}), the loss structure
in velocity space is not a cylinder as in the case of direct capture
[see eq.\ (\ref{eqn:cylind})]. Rather, this structure is fatter near
$v_r\sim 0$ and narrower near $v_r\sim \pm v_\esc$.

\section{Observable Consequences}

  The large number of black holes that move to the center by dynamical
friction have several observable effects which we now discuss.

  The most important effect is that the stars that formerly resided in
the region that is now occupied by the cluster of black holes are
ejected into orbits at larger radius. Therefore, the old
stellar population of mass $m$ should have a very large core radius,
given by $r_{cs} = r_\bh (m_\bh/m)^{2/7} \sim 1 - 2\, \pc$.
This core radius ($\sim 40''$) is much larger than the value
expected from stellar collisions alone, which would only produce a core
at the radius $\sim 0.03\, \pc$ where the orbital velocities are
comparable to the escape velocities from the stellar surfaces.
The most straightforward test of our model is therefore to measure the
distribution of low-mass stars and find out if this very large core
indeed exists. 

  In fact, the bright ($K<15$) stars in the inner galaxy exhibit
a power-law profile ($\alpha\sim 7/4$) all the way to $r\sim 0.1\,\pc$
where a core sets in (Genzel et al.\ 2000; Schmitt 1995).  However, 
this is not necessarily inconsistent with a core in the old population:
most of the observed stars could well be young.
Indeed, Genzel et al.\ (2000) have found that the stars they observed
very near to the Galactic center are young supergiants, which must have 
formed in a recent ($\sim 10\,$Myr) burst of star formation.  If such
bursts of star formation have occurred repeatedly, then we would expect that 
the inner region $r<r_\bh\sim 0.7\,\pc$ should contain the remains of
the bursts from the last $t_r({r_\bh})\sim 2\,$Gyr, but not earlier.
Blum, Sellgren, \& DePoy (1996a) found that the luminosity function in the
inner $2'$ (5 pc) extends to several magnitudes brighter than the one in 
Baade's
Window and inferred that young stars must be present in significant numbers.
Blum, Sellgren, \& DePoy (1996b) obtained spectra of 19 of these brighter
stars.  Of those they could age-date, most were younger than 200 Myr.

	The intermediate-age ($\sim 2\,$Gyr) and older stars ejected from
the central cluster could be expected to be found up to a few $r_0$ from
Sgr A* on ``Oort Cloud'' like orbits.  That is, a star from this
population would
keep getting jolted by the black holes to a more and more eccentric orbit
until diffusive encounters near apocenter drove it into an orbit with a
pericenter just beyond $r_\bh$.  Narayanan, Gould, \& Depoy (1996) list
16 presumably intermediate-age stars $(K_0<5)$ at projected radii of
8 to 20 pc.  If these come primarily from the central pc, they should
be preferentially on radial orbits.  Of course, there has been recent
star formation outside the central pc as well.  For example,
Blum et al.\ (1996b) estimate the ages IRS 24 and IRS 23 at $\sim 100\,$Myr
and $\sim 200\,$Myr respectively, and these both lie at a projected
distance of 1.7 pc.
Hence, not all the Narayanan et al.\ (1996) stars are necessarily ejected.

  Another interesting consequence of the ejection of low-mass stars from the
center is that the capture of ordinary stars by supermassive black holes
could be much rarer than commonly believed. 
These captures should lead to bright optical
flares (e.g., Rees 1988; Ulmer 1999), which could be found in supernova
searches. A cluster of black holes should have formed around all the
supermassive black holes in galactic centers by the dynamical friction
process described here, and these clusters
would reduce by a large factor the rate at
which stars can come close enough to the black hole to be tidally
disrupted.  Of course, since there is a central density cusp containing
at least young stars around Sgr A*, there will be some tidal captures,
but the absence of old stars should reduce the number of expected flares.
In addition, for galaxies such as ellipticals that are poor in neutral
gas, one would not expect continuous star formation
near the central black holes. Consequently, after the black hole cluster
had ejected all the old stars, no young stars would form to replace them.
For these galaxies, flares from stellar captures could be extremely rare.

  We mention also microlensing of a background bulge star as another
potentially observable effect, although requiring a large improvement in
sensitivity and resolution of infrared imaging over our current
capabilities. If a bulge star at a distance $2\, \kpc$ behind the center
could be observed, the angular Einstein radius of 
Sgr A* would be $b = 0.\hskip-2pt ''8$, corresponding to a linear
size of $r_E=0.03\, \pc$. The two images of the star could therefore be
comfortably resolved. The star would typically take several hundred
years to complete the ``microlensing event'' by Sgr A*.
These two images would then have a microlensing optical depth to be
lensed by one of the cluster black holes,
$\tau \sim (N_\bh m_\bh/M_{BH})(r_{E}/r_{bh})^{5/4}\, A\sim 10^{-3} A$,
where $A$ is the magnification of the image. Imaging down to $K=21$,
one should on average find about one background star at $2\, \kpc$
within an Einstein radius of Sgr A*, and about 100 similar stars near the
core radius of $\sim 1\,\pc$ (for which the Einstein radius of Sgr A*
would be only $\sim 0.\hskip-2pt ''02$). Of course, the identification
of the two images of a star lensed by Sgr A* would be of enormous
interest by itself; despite the large number of orbiting black holes in
the cluster, the expected rate of the ``planet-like'' events is still
low, and several lensed stars would need to be identified to have a good
chance of detecting the short events over a period of $\sim 10$ years.

%  Finally, the gravitational waves of the black holes emitted in their
%peribothron passages might be detectable by LISA. This subject will be
%discussed in a separate paper.

\acknowledgements

Work by AG was supported in part by grant AST 97-27520 from the NSF.
We are happy to acknowledge discussions with Martin J. Rees.

\appendix
\renewcommand{\theequation}{\thesection\arabic{equation}}
\section{Diffusion Tensor in a Kepler Potential}

Here, we calculate the diffusion tensor $\kappa$ for the general case of
a $\rho\propto r^{-\alpha}$ density profile in a Kepler potential.
assuming that the velocity distribution is isotropic.  Recall that the
phase space distribution is given by equations (\ref{eqn:aone}) and
(\ref{eqn:atwo}). 
%\begin{equation}
%f_0({\bf u,r}) = g\biggl({u^2\over v_\esc^2}\biggr)h(r),\qquad 
%g(x) \equiv (1 - x)^{\alpha-3/2}
%\Theta(1-x),
%\label{eqn:aone}
%\end{equation}
%\begin{equation} 
%h(r)\equiv
%{(3-\alpha)\alpha!\over 8\pi^2(\alpha-3/2)!(1/2)!}
%{N_\bh\over(2 G M_\cbh r_\bh)^{3/2}}
%\biggl({r\over r_\bh}\biggr)^{3/2 - \alpha}
%\label{eqn:atwo}
%\end{equation}
%where $v_\esc(r)$ is the local escape velocity, $N_\bh$ is the total number
%of particles within a radius $r_\bh$, $M_\cbh$ is the central mass, and
%$\Theta$ is a step function.  
We begin our evaluation at ${\bf v}=0$ where the diffusion tensor, $\kappa_0$,
is isotropic.  A single encounter with another black hole at speed $u$
changes the velocity by $\Delta v = 2 G m_\bh/(b u)$.  Hence the
time averaged growth of $\langle v^2\rangle$ is
$$ {d  \langle v^2\rangle\over d t} =
\int 2\pi b d b\int d^3 u \biggl({2 G m_\bh\over b u}\biggr)^2 u f_0(u,r) $$
\begin{equation}
= {2(3-\alpha)\alpha!\over (\alpha-1/2)!(1/2)!}
{N_\bh(G m_\bh)^2 v_\esc^2\over(2 G M_\cbh r_\bh)^{3/2}}
\biggl({r\over r_\bh}\biggr)^{3/2 - \alpha}\ln \Lambda_1 ~,
\label{eqn:aseven}
\end{equation}
where $\Lambda_1$ is the ratio of maximum to minimum impact parameters,
which we evaluate below.

	If the diffusion equation is written (for isotropic $\kappa_0$) as
$\kappa_0\nabla_v^2 f = \partial f/\partial t$, then 
\begin{equation}
\kappa_0 = {1\over 6}{d\langle v^2\rangle\over d t} ~,
\label{eqn:aeight}
\end{equation}
which combined with equation (\ref{eqn:aseven}) gives $\kappa_0$.

	Next, we evalute $\kappa_\perp(v)$ and  $\kappa_r(v)$ for the
special case of $\alpha=3/2$ for which the phase-space density (eqs.\
[\ref{eqn:aone}] and [\ref{eqn:atwo}]) is independent of both radius and
velocity and consequently the problem is tractable analytically.  Up to
an overall normalization $\gamma$, the transverse velocity diffusion is
given by
\begin{equation}
{d  \langle v_\perp^2\rangle\over d t} =
{\gamma\over 2} \int_{-v_\esc}^{v_\esc} d u_r
\int_{\theta=0}^{\theta_\max(u_r)}\biggl({1+\cos^2\theta\over 2}\biggr)
{ (u_r + v)^2 d\tan^2\theta\over |u_r+v|\sec\theta},
\label{eqn:btwo}
\end{equation}
where $\cos\theta_\max(u_r)\equiv |u_r + v|/(v_\esc^2 + v^2 + 2 u_r v)^{1/2}$.
For the total velocity diffusion ${d  \langle v_\perp^2\rangle/d t}$,
we find a similar expression, but with $(1+\cos^2\theta)/2\rightarrow 1$.
We evaluate these expressions and find
\begin{equation}
{d  \langle v^2\rangle\over d t}= \gamma v_\esc^2\biggl(1 - {1\over 3}
{v^2\over v_\esc^2}\biggr),\qquad
{d  \langle v_\perp^2\rangle\over d t}= {2\over 3}
\gamma v_\esc^2\biggl(1 - {1\over 5}
{v^2\over v_\esc^2}\biggr).
\label{eqn:bthree}
\end{equation}
We therefore conclude that 
$\kappa_\perp(v)/\kappa_0 = 1-0.2(v/v_\esc)^2$ and
$\kappa_r(v)/\kappa_0 = 1-0.6(v/v_\esc)^2$.
These results apply to $\alpha=3/2$, but we adopt them for $\alpha=7/4$
as well.  If the true coefficient for $\alpha=7/4$ is $0.2(1+\epsilon)$
rather than 0.2, then this introduces an error into the capture formula
(\ref{eqn:capthree}) of only $(1/6)\epsilon$.  Since $\epsilon$ is likely
to be small, this correction is negligible.

	We now justify two other approximations made in the capture
calculation in \S\ 3.  First, in equation (\ref{eqn:diffsolv}), we
effectively considered the diffusion coefficient as fixed on {\it slices}
of the velocity sphere perpendicular to the cylinder and passing through
$v_r$.  In fact, it is fixed on spherical shells of constant speed.  This
``plane parallel'' approximation is justified by two related considerations.
First, $\kappa$ varies very slowing, only by a factor $1.25$ over the
entire velocity sphere.  Second, most of the depleted region of velocity
space is relatively near the cylinder, so that the error in $\kappa$ made
by the plane-parallel approximation is extremely small.

	Next, the boundary condition for the solution to the diffusion
equation (\ref{eqn:diffsolv}) sets the phase-space density $f$ at the
edge of the velocity sphere $(u=v_\esc)$ equal to the unperturbed density
$f_0$ at $u=v_r$.  Strictly speaking, if the boundary condition is set at
$(u=v_\esc)$, then one should use $f_0(v_\esc)$.  However, for the $\alpha=7/4$
profile, this is zero.  (Our approximation would be exact for $\alpha=3/2$).
The basis for our approximation is that the density returns essentially
to $f_0$ long before the edge of the velocity sphere, at which point $f_0$
is not much different from $f_0(v_r)$.

	Finally, we evaluate $\Lambda_1=b_\max/b_\min$, the ratio of the
maximum to minimum impact parameters.  This can also be written
$\Lambda_1 = \epsilon_{\rm max}/\epsilon_\min$, where 
$(\epsilon_\min v_\esc,\epsilon_\max v_\esc)$ is the range of impulses
relevant to the problem at hand.  For general relaxation
$\epsilon_{\rm max}=1$ and $\epsilon_\min\sim 2Gm_\bh/(r v_\esc^2)=
m_\bh/M_\cbh$.  However, while the harder
scatters all contribute to the {\it overall} relaxation, they do not 
contribute to diffusion into the loss cone because the black hole
simply ``jumps over'' the capture cylinder.  Unfortunately, it is not
trivial to identify exactly what the largest allowed jumps should be:
while jumps larger than $(q_\min/r)^{1/2}v_\esc$ do not directly lead
to capture, they do help maintain the overall velocity profile given
by equation (\ref{eqn:diffsolv}).  Note that the profile differs significantly
from the background for many $e$-foldings.  If the jumps larger than
$(q_\min/r)^{1/2}$ did not contribute at all, then 
$\Lambda_1=(q_\min/r)^{1/2}\Lambda$.  In view of the distribution's 
slow approach to the background level near the capture cylinder, we adopt
$\Lambda_1 = \Lambda(q_\min/r)^{1/4} = 9.7$.


\begin{references}

\reference{} Bahcall, J.N., \& Wolf, R.A.\ 1976, \apj, 209, 214
\reference{} Bailyn, C.D., Jain, R.K., Coppi, P., \& Orosz, J.A.\ 1998, 
\apj, 499, 367
\reference{} Blum, R.D., Sellgren, K., \& DePoy, D.L.\ 1996a, \apj, 470, 864
\reference{} Blum, R.D., Sellgren, K., \& DePoy, D.L.\ 1996b, \aj, 112, 1988
\reference{} Eckart, A., \&  Genzel, R.\ 1997, \mnras, 284, 576
\reference{} Frank, J., \& Rees, M. J. 1976, \mnras, 176, 633
\reference{} Genzel, R., Pichon, C., Eckart, A., Gerhard, O.E., \& Ott, T.\
2000, \mnras, submtited
\reference{} Gould, A., Bahcall, J.N., \& Flynn, C.\ 1997, \apj, 482, 913
\reference{} Han, C., \& Gould, A.\ 1996, \apj, 467, 540
\reference{} Lightman, A. P., \& Shapiro, S. L. 1977, \apj, 211, 244
\reference{} Narayanan, V.K., Gould, A., \& Depoy, D.L.\ 1996, \apj, 472, 183
\reference{} Quinlan, G.D., Hernquist, L., \& Sigurdsson, S.\ 1995, \apj, 440,
554
\reference{} Rees, M. J. 1988, Nat, 333, 523
\reference{} Richstone, D., et al.\ 1998, Nature, 395, A14
\reference{} Schmitt, J.\ 1995, Thesis (Munich: Ludwig-Maximilian University)
\reference{} Sigurdsson, S., \& Rees, M. J. 1997, \mnras, 284, 318
\reference{} Ulmer, A. 1999, ApJ, 514, 180
\reference{} Zoccali, M., S., Cassisi, S., Frogel, J.A., Gould, A.,
Ortolani, S., Renzini, A.,  Rich, R.M. 1999, \&
Stephens, A.\ 2000, \apj, in press (astro-ph/9906452)

\end{references}
\end{document}